\documentclass[10pt,conference]{IEEEtran}
\IEEEoverridecommandlockouts
\usepackage{cite}
\usepackage{amsmath,amssymb,amsfonts}
\usepackage{algorithmic}
\usepackage{graphicx}
\usepackage{textcomp}
\usepackage{xcolor}%
\usepackage{booktabs} 
\usepackage{multirow}
\usepackage{xspace}
\usepackage{balance}
\usepackage{url}
\usepackage{caption}
\usepackage{subcaption}
\usepackage[ruled,linesnumbered]{algorithm2e}
\usepackage[capitalize,noabbrev]{cleveref}
\usepackage[normalem]{ulem}
\usepackage{listings}

\newcommand{\name}{\textit{TRaf}\xspace} % THE NAME OF THE APPROACH!
\begin{document}

%\title{Conference Paper Title*\\
\title{
    
    %Automated Time-based Treatment of Asynchronous Wait Flaky Tests in Web Testing
    %\name: An Automated Time-based Repair for Asynchronous Wait Flaky Tests in Web Testing
    Time-based Repair for Asynchronous Wait Flaky Tests in Web Testing
}
% \author{\IEEEauthorblockN{Anonymous Author(s)}
% }

\author{\IEEEauthorblockN{Yu Pei}
\IEEEauthorblockA{SnT, University of Luxembourg \\
Luxembourg\\
yu.pei@uni.lu}
\and
\IEEEauthorblockN{Jeongju Sohn}
\IEEEauthorblockA{SnT, University of Luxembourg \\
Luxembourg \\
jeongju.sohn@uni.lu}
\and
\IEEEauthorblockN{Sarra Habchi}
\IEEEauthorblockA{Ubisoft \\
Canada\\
sarra.habchi@ubisoft.com}
\and
\IEEEauthorblockN{Mike Papadakis}
\IEEEauthorblockA{SnT, University of Luxembourg\\
Luxembourg \\
michail.papadakis@uni.lu}
}

\maketitle
\IEEEpeerreviewmaketitle
\begin{abstract}
Asynchronous waits are one of the most prevalent root causes of flaky tests and a major time-influential factor of web application testing. To investigate the characteristics of asynchronous wait flaky tests and their fixes in web testing, we build a dataset of 49 reproducible flaky tests, from 26 open-source projects, caused by asynchronous waits, along with their corresponding developer-written fixes. Our study of these flaky tests reveals that in approximately 63\% of them (31 out of 49), developers addressed Asynchronous Wait flaky tests by adapting the wait time, even for cases where the root causes lie elsewhere. 
Based on this finding, we propose \name, an automated time-based repair method for asynchronous wait flaky tests in web applications. \name tackles the flakiness issues by suggesting a proper waiting time for each asynchronous call in a web application, using code similarity and past change history. The core insight is that as developers often make similar mistakes more than once, hints for the efficient wait time exist in the current or past codebase. 
Our analysis shows that \name can suggest a shorter wait time to resolve the test flakiness compared to developer-written fixes, reducing the test execution time by 11.1\%. With additional dynamic tuning of the new wait time, \name further reduces the execution time by 20.2\%.  
\end{abstract}

%\IEEEpeerreviewmaketitle


\section{Introduction}\label{sec:introduciton}

Regression testing ensures the system under test (SUT) behaves as expected when software is changed: a test failure implies that the latest changes introduce faults into the SUT~\cite{hilton2016usage,Yoo2010fk,Labuschagne2017FSE}. In practice, however, some tests fail intermittently during regression testing, even in the absence of fault-introducing changes and sometimes on the same version of the system. These tests, called flaky tests, force developers to waste their effort debugging them, only to discover later that they are false alarms. This causes developers to overlook even the true failure alarms due to their past experiences with flaky tests.

Flaky tests are quite common in modern software systems~\cite{LeongICSE2019google,Labuschagne2017FSE,Luo2014,Romano2021ICSE,Gruber2021}. Labuschagne et al. showed that 13\% of failed test executions in open-source projects using Travis CI were flaky~\cite{Labuschagne2017FSE}. At Google, despite the persistent effort to remove flaky tests, around 16\% of the tests still exhibit a certain flakiness degree, and 84\% of transitions between Pass and Fail were caused by flaky tests, indicating the importance of flakiness issues in CI environment~\cite{Micco2017,LeongICSE2019google}. While flaky tests appear inevitable, manually addressing them is infeasible due to their sheer number and ambiguous nature. Hence, many studies have focused on automatically resolving them, either entirely removing root causes or reducing the chance of being flaky~\cite{shi2019ifixflakies,Lam2020a,Li2022icseODRepair,Dutta2021fseFLEX}. As various factors (i.e., root causes) can incur flaky tests, each requiring a different approach, the proposed repair techniques target a single category of flaky tests caused by the same root cause. 

Asynchronous call/wait is one of the most common causes of flaky tests~\cite{Luo2014,Romano2021ICSE,Lam2020a}. Nevertheless, compared to other categories of flakiness, such as order-dependency, not much work has been done to repair it automatically. 
Recently, Lam et al. presented FaTB, short for Flakiness and Time Balancer, that refines developer-written fixes for Async Wait flaky tests, i.e., flaky tests caused by asynchronous calls/waits, to reduce the execution time while retaining the same flake rate as the original fixes~\cite{Lam2020a}. 
Their study on six large-scale proprietary projects in Microsoft showed the prevalence of Async Wait flaky tests and developers' preference to handle them simply by increasing the wait time. From this, the authors performed a small study on the effectiveness of FaTB in refining the wait time in developer fixes to balance between the decreased flake rate and increased test execution time; through this study, they demonstrated the potential benefits of collective reduction in test execution time by using more efficient time values for waiting in the Async Wait tests. 
Prior studies in web testing revealed that web developers often employ large wait times to handle the flakiness caused by asynchronous calls~\cite{presler2019wait,olianas2022sleepreplacer}. %, such as DOM or page condition issues, 
This paper also aims to address Async Wait flakiness efficiently, i.e., minimizing the cost in test execution time, but specifically in web front-end testing and, more importantly, without relying on user-related inputs.

Testing in web environments is particularly challenging as testing goes beyond unit and integration tests and includes front-end testing. Web front-end testing often involves user interaction through the graphical user interface (GUI). Due to the complexity of user interactions, network instability, and the inherently asynchronous nature of web applications, web front-end testing is more subject to flaky tests than traditional unit testing. Yet, the research community paid less attention to flaky tests in web front-end testing than traditional unit tests. 

Given that our focus is limited to Async Wait flaky tests, we conduct a study on the practice of developers to resolve Async Wait flaky tests in this specific test domain. We study 49 Async wait flakiness from 26 web projects, examining the flakiness in test code and their corresponding developer-written repairs. From this study, we find that in approximately 63\% of the cases (31 out of 49), developers repaired Async Wait for flaky tests by adding or increasing the time to wait, concurring with the previous findings on the fixes~\cite{Luo2014,Romano2021ICSE,Hashemi2022icsme,Lam2020a, malm2020automated, Parry2021}.

Based on the finding of the preliminary study, we propose \name, short for Time-based Repair of Async Wait Flaky tests, a new repair framework that automatically addresses Async Wait flaky tests in web front-end testing. \name uses the code similarity and past change history to find the likely wait time to repair a given flaky test. Our basic idea is that fixing ingredients for the repair, in our case, the time values, exist in the past or current codebase. 
%\name is static by default and thus does not require any test rerun until validating the generated repair with the suggested time value, maximizing its usability; 
\name statically suggests a set of likely new wait times and does not involve any test run until validating the generated repair with the suggested time value, maximizing its usability; 
it allows users to dynamically tune the suggested wait time further, searching for the optimal time between the original flaky time and the newly proposed one. 

Experimental results with 31 Async Wait tests repaired by adding or increasing the timeout (out of 49 studied tests) show that \name can propose an efficient wait time that reduces the test execution time by an average of 11.1\% and of 20.2\%, without and with the dynamic optimization, compared to developer-fixed tests, while still relieving the test flakiness. 

In summary, the main contributions of this paper are:

\begin{itemize}
    \item We derive a reproducible dataset of 26 JavaScript open-source web projects with 49 tests exhibiting Async Wait flakiness.

    \item We investigate the fixing strategies behind tests exhibiting Async Wait flakiness in web front-end testing and confirm that adding or increasing wait time, employed in 31 out of 49 studied tests, is the most common fixing practice for Async Wait flaky tests. 

    \item We propose \name, a new repair technique for Async Wait flaky tests in web front-end testing. \name efficiently finds an adequate wait time using the code similarity and version history of the code under testing. 

    \item Experimental results with 31 flaky tests show that \name can reduce the test execution time by 11.1\% on average than the original developer fixes while still addressing the flakiness in all studied tests; with additional dynamic optimization, it can reduce the execution time by 20.2\%.
    
\end{itemize}

\section{Async Wait Flaky Tests in Web Front-End Testing}\label{sec:async_wait_flakiness}

Continuity of program execution during a call process does not guarantee that the result will be accessible when needed. For instance, a program may request a resource that is not fully rendered or attempt to process data yet to be loaded due to network delays. A common approach to deal with such flakiness caused by an asynchronous call is to wait for a certain amount of time to allow the call or action to complete before executing subsequent actions. However, because of the inherently unpredictable nature of web testing and the diversity of web resources, the web access load is difficult to estimate; each asynchronous call may incur a different timeout. Thus, there is no guarantee that the call will complete within a pre-defined time. That is where the asynchronous waiting flaky behavior originates, causing a test to pass and fail non-deterministically.

Async Wait flaky test failures in web application testing typically occur in the following scenario: a test interacts with the Document Object Model (DOM), waits for resources to be rendered/loaded, and performs subsequent actions, such as transition or animation~\cite{Romano2021ICSE}. Hence, DOM and time are the primary concerns of developers and testers for web front-end testing since they directly influence how web pages display and load. In light of this, we investigate the role of the time and DOM elements in triggering and resolving Async Wait flaky tests in our preliminary study detailed in Section~\ref{sec:subject_study}.

\section{Preliminary Study}\label{sec:subject_study}

This section explores how developers introduce and handle Async Wait flaky tests, mostly related to DOM and time issues in web front-end testing. This preliminary study aims to gain insight into how developers treat, i.e., fix, flaky tests.

\subsection{Data Collection}\label{sub:data_collection}

We focus on flaky tests caused by the Asynchronous Wait mechanism (i.e., Async Wait flaky tests) in web applications since they typically contain various asynchronous events, such as user interaction with the web interface~\cite{Romano2021ICSE}.
%such as those interacting with network resources~\cite{Romano2021ICSE}.
To collect a set of Async Wait flaky tests, we followed a procedure similar to that of Gruber et al., searching for commits that fix flakiness in open-source projects on GitHub with certain keywords~\cite{Gruber2021}. We then rerun the tests in those commits a fixed number of times and check whether there is any change in the test state among the runs. 

Specifically, we first looked for the keywords \textit{"e2e"}, \textit{"flaky"}, \textit{"flakiness"}, and \textit{"Intermittent"} in commits. The GitHub search identified over 800 distinct repositories related to these topics. Our exploration focused on code repositories and tests written in JavaScript, one of the most common web development languages, to narrow our scope of web front-end testing. 
We further filtered out the obtained commits using the keywords \textit{"async"}, \textit{"wait"}, \textit{"timeout"}, \textit{"DOM"}, and \textit{"delay"} to ensure they relate to Async Wait flakiness. We then manually identified the commits modifying the test code and isolated those that are potentially related to fixing Async Wait flakiness. Thus, we found roughly 300 projects with commits potentially fixing Async Wait flakiness. 

To obtain a reproducible set of async wait flaky tests, we cloned the projects from GitHub and reran related tests modified in the resulting commits from the previous step. To capture the flakiness, we executed each suspicious candidate test 100 times and labeled it as flaky if its pass/fail outcome changed in any of its executions; we saved the error messages for later analysis. 
Most of the 300 projects with flaky commits could not be executed in our local environment due to issues with dependencies and inconsistencies between operating systems. For many of those that successfully recreated the testing environment, we failed to reproduce the flakiness reported by developers in 100 test reruns; the failures might be attributed to insufficient test reruns or to some unresolved inconsistencies between test environments. We ended up with 26 projects, as shown in \cref{tab:projects}. With the manual inspection of test outcomes and corresponding error messages, we eventually obtained 49 reproducible flaky tests among 1,869 suspicious flakiness-related commits across 26 web application projects. \cref{tab:projects} shows the distribution of the number of tests per project. 

\begin{table}[ht] 
  \caption{Subjects. \#$_{all}$ and \#$^{related}_{flake}$ are the number of total commits and flakiness-inducing commits. \#$^{rerun}_{flake}$, \#$^{dom}_{cause/fix}$ and \#$^{time}_{cause/fix}$ are the number of reproduced flaky tests and the number of these flaky caused/fixed by DOM-based and Time-based factors.}\label{tab:projects}
  %\vspace{-1.0em}
  \scalebox{0.7}{
  \begin{tabular}{l|rrr|rr|rr}
      \toprule
       & \multicolumn{3}{c|}{Commits} & \multicolumn{2}{c|}{Cause} & \multicolumn{2}{c}{Repair} \\
      project & \#$_{all}$ & \#$^{related}_{flake}$ & \#$^{rerun}_{flake}$ & \#$^{dom}_{cause}$ & \#$^{time}_{cause}$ &  \#$^{dom}_{fix}$  &  \#$^{time}_{fix}$ (\#$^{dom}_{cause}$/\#$^{time}_{cause}$)  \\
      
      \midrule
      jbrowse & 8,434 & 79 & 6 & 6 & 0 & 0 & 6 (6/0) \\
      elem-ing & 579 & 75 & 5 & 5 & 0 & 5 & 0 (0/0)  \\
      uploady & 668 & 21 & 4 & 0 & 4 & 0 & 4 (0/4) \\
      flashmap & 7,325 & 105 & 4 & 4 & 0 & 0 & 4 (4/0) \\
      react-rxjs & 387 & 12 & 3 & 0 & 3 & 0 & 3 (0/3) \\
      puppeteer & 516 & 77 & 2 & 0 & 2 & 0 & 2 (0/2) \\
      predictive & 1,198 & 22 & 2 & 0 & 2 & 0 & 2 (0/2) \\
      keptn & 8,273 & 189 & 2 & 2 & 0 & 1 & 1 (1/0) \\
      dom-testing & 171 & 24 & 2 & 2 & 0 & 2 & 0 (0/0) \\
      preact-dev & 1,416 & 43 & 2 & 2 & 0 & 1 & 1 (1/0) \\
      coinscan & 184 & 11 & 2 & 2 & 0 & 1 & 1 (1/0) \\
      others\footnotemark[1] & 81,467 & 1,211 & 1 (15) & 1 (11) & 1 (4) & 8 & 7 (3/4)\\
      \midrule
      total & 110,618 & 1,869 & 49 & 34 & 15 & 18 & 31 (16/15) \\
      \bottomrule
  \end{tabular}
  }
\end{table} 
\footnotetext[1]{%The following projects contain one flaky test per project: 
material-ui, dotcom-rendering, codacollection, elm-select, eui, edelweiss-ui, racp, wonder-blocks, client, liquid, shopify-theme-inspector, azure2jira, beacons, js-libp2p, cordless.
}
 
We categorized the collected Async Wait flaky test by inspecting the related commits, test code, and error messages; this manual categorization took approximately 40 working days to complete.
We call tests \textit{DOM-based} if the flaky synchronization point contains an explicit call to a DOM element, such as \texttt{waitForElements} or \texttt{waitForBeTrue}. If a test is flaky on a call to wait for an explicit amount of time, we refer to it as \textit{Time-based}. \cref{tab:projects} presents 34 DOM-based and 15 Time-based tests in our dataset regarding their cause, i.e., the type of asynchronous calls. 

\subsection{Categories of Developer fixes}\label{sub:categories_of_developer_fixes}

We noted that developers typically increase or add a timeout to the test or modify it to wait for a specific DOM element to be available to reduce flakiness. Hence, we define two repair strategies for Async Wait flakiness, similar to what we did for the root causes: \textit{Time-based} and \textit{DOM-based}. 

\subsubsection{DOM-based fixes}\label{subsub:DOM_based_fixes}

\begin{figure}[h]
  \centering
  \includegraphics[width=0.88\linewidth]{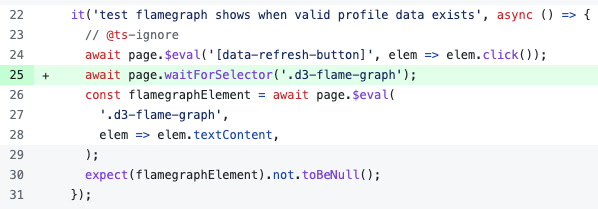}
  \vspace{-0.2em}
  \caption{DOM-based fixes example.}
  \label{fig:dom-based}
\end{figure}

Implementing a \texttt{waitFor} with a DOM condition in a test is a common way to fix async wait flakiness. This type of fix, i.e., DOM-based, covers different DOM-related conditions, such as presence, visibility, and availability. In summary, DOM-based fixes repair the synchronization point that depends on the rendering state of a specific DOM element. 
The main objective of the DOM-based fixes is to guarantee that the test will be executed after the web page elements complete rendering attributes, states, or data. 
One such example is shown in \cref{fig:dom-based}. 
The code snippet is from an Async Wait flaky test we examined in the Shopify-theme-inspector; by rerunning this test, we got an error message, \textit{"Failed to find an element matching the selector d3-flame-graph"}. The developer repairs this flaky test by adding \texttt{waitForSelector} condition to fully load the 'd3-flame-graph' element before the assertion in Line 25. 
In our dataset, 18 of 49 tests were fixed using the DOM-based fix strategy, as shown in column \#$^{dom}_{fix}$ in \cref{tab:projects}.

\subsubsection{Time-based fixes}

\begin{figure}[h]
  \centering
  \includegraphics[width=0.88\linewidth]{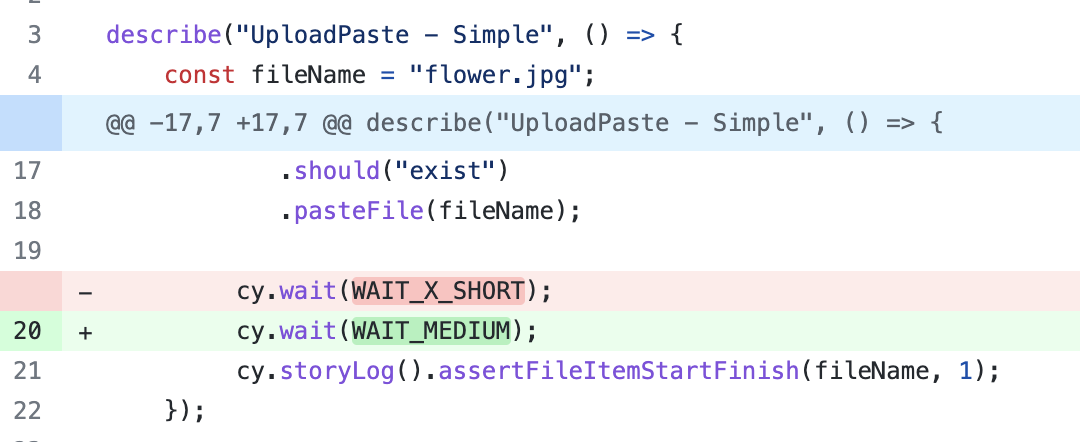}
  \vspace{-0.3em}
  \caption{Time-based fixes example.}
  \label{fig:time-based}
\end{figure}

The most common way to resolve Async Wait flakiness is by increasing or adding the wait time. The Time-based fixes address the flakiness by waiting for an explicit amount of time at the synchronization point. 
\cref{fig:time-based} presents the code snippet of such a Time-based fix from React-uploady. 
%One such example is shown in Figure~\ref{fig:time-based}. where developers repair the flaky test by increasing the wait time from \texttt{WAIT\_X\_SHORT(250 ms)} to \texttt{WAIT\_MEDIUM(1500 ms)} to ensure the completion of the file paste process before the assertion at line 20. \CHECKMEYU{
When we reran this test, we received an error message, \textit{"AssertionError in storyLog: expect matching ITEM\_FINISH for ID: batch-1.item-1"}. Developers addressed this flakiness failure by increasing the wait time from \texttt{WAIT\_X\_SHORT(250 ms)} to \texttt{WAIT\_MEDIUM(1500 ms)} to ensure the completion of the file paste process before the assertion \texttt{assertFileItemStartFinish} at line 20.
%Developers repaired this flakiness failure by increasing the wait time before the \texttt{assertFileItemStartFinish} statement.

In our dataset, as \cref{tab:projects} shows, 31 out of 49 tests used the time-based repair strategy, accounting for the majority, with around 51\% (16 out of 31) of tests that were flaky for DOM-based causes, as shown in column \#$^{time}_{fix}$ in \cref{tab:projects}. Most of the tests being flaky by Time-based reasons were fixed by increasing the wait time (i.e., 13 out of 15 cases).\footnote{There are two cases where the tests flaky for Time-based reason, based on the source code, were resolved by simply adding a \texttt{waitTime()} statement.} We speculate that this dominance of the time-based repair strategy is related to the difficulty of identifying for which element to wait; depending on the design and complexity of the system, identifying the right component to wait for may not be straightforward. 

Regarding this difficulty of fixing Async Wait flaky tests, we further look into the time values in flaky tests and those chosen by developers to address the flakiness, specifically for flaky tests incurred and resolved by the Time-based cause and strategy (i.e., \#$^{time}_{cause/fix}$ in the last column of \cref{tab:projects}). Overall, the wait time in flaky tests ranges from 0 to 1500 ms with an average of 466 ms;\footnote{wait 0: a task is executed at the earliest available idle time of the main thread, it is due to JavaScript in a browser executing on a single thread.} the time in the developer fixes is between 50 and 30000 ms with an average of 1282 ms.

% modified\CHECKME{Time-based fix, 53\% of dom cause, 47\% of time cause } 

%Previous work of Lam et al. on six large-scale proprietary projects at Microsoft~\cite{Lam2020a} shows that the majority of developer-written fixes for Async Wait flaky tests involve increasing wait time; based on the additional investigation, they conclude that developers prefer to handle async wait tests 

The large gap between the time values in flaky tests and in developer fixes implies two conflicting goals during web front-end testing: 1) to reduce execution time and 2) to alleviate test flakiness. Our study shows that developers often set a short wait time to reduce test execution costs and switch to a longer time to alleviate flakiness. A recent study by Lam et al.~\cite{Lam2020a} showed that developers often increase wait times to fix Async Wait flaky tests and that these time values can be refined to reduce test effort, i.e., execution time.
This paper aims to propose an efficient way to repair Async Wait flaky tests by finding a fair wait time to reduce flakiness and execution time.

\section{Automated Repair of Async Wait Flaky Tests}\label{sec:automated_repair}

\begin{figure*}[ht!]
\centering
\includegraphics[width=0.6\textwidth, trim = 4mm 0mm 0mm 16mm, clip]{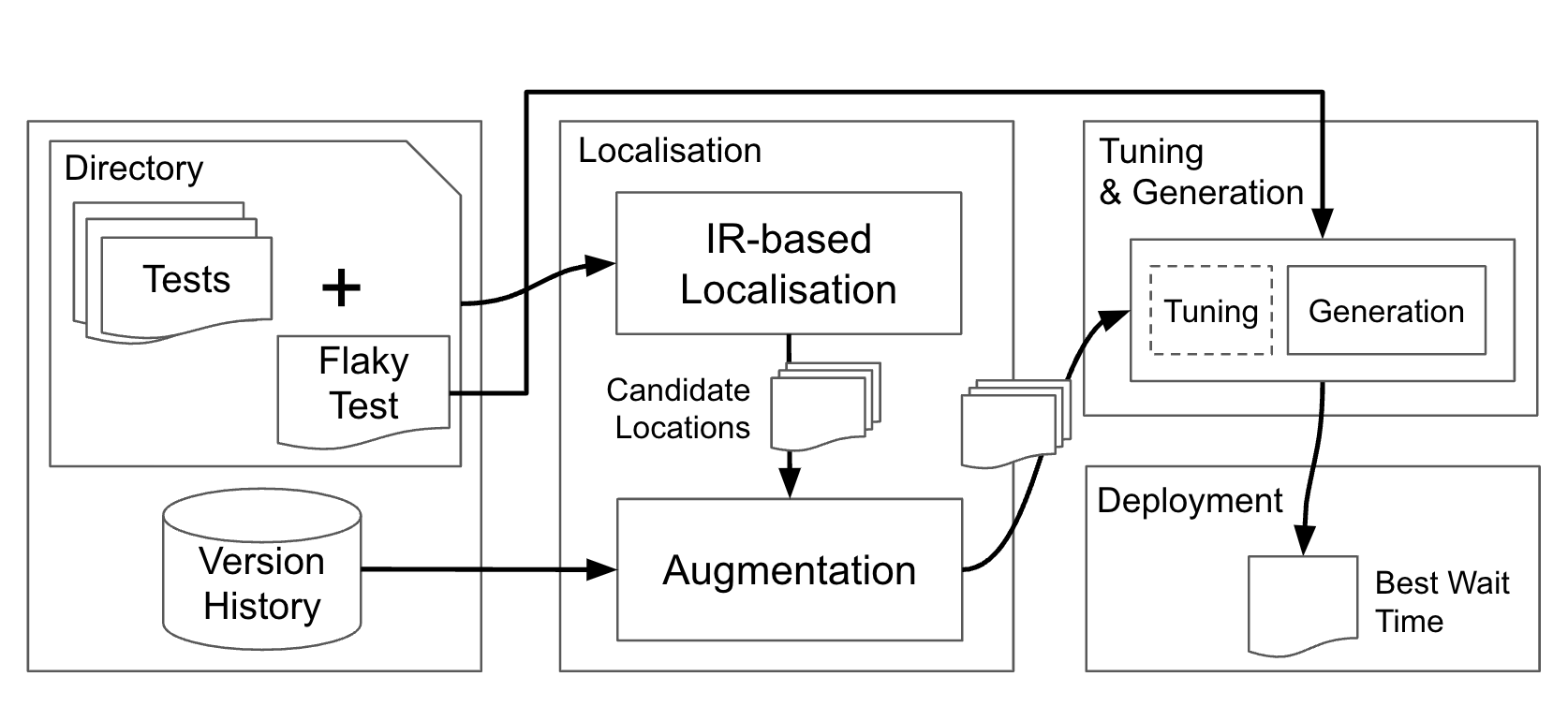}
\vspace{-1.0em}
\caption{
Overall Architecture of \name. The dotted borderline means users can skip the Tuning. 
}\label{fig:archiecture}
\end{figure*}
%\vspace{-0.5em}

Based on the findings of our preliminary study, we propose \name, short for Time-based Repair of Async wait Flaky tests, a new automated time-based repair framework for Async Wait flaky tests in web front-end testing. 
\cref{fig:archiecture} shows the overall architecture of \name. \name consists of two main phases: 1) localization and 2) tuning and generation. 
In the localization phase, \name uses an information retrieval (IR)-based approach to identify code lines to extract candidates of the new wait time for a given flaky test. It then augments this candidate set by analyzing the version history of the code under inspection. 

In the tuning and generation phase, \name first ranks the candidate lines based on their likelihood to contain an adequate wait time and extracts new time values sequentially from the top. 
It then uses the extracted values to generate patches for the flaky test or goes through an extra step of dynamic tuning of these time values before using them for repair.

\subsection{Localization}\label{sub:localization}
 
In order to automatically repair a test failure, including our target of Async Wait test flakiness, we need ingredients to compose a fix-patch in addition to fix locations and operators. The plastic surgery hypothesis is one of the most adopted strategies to get fixing ingredients in the Automatic Program Repair (APR) domain~\cite{Goues2012TSE,Matias2014Icse,Wen2018:cse,Barr:2014sf,TBar2019Issta}. The underlying intuition is that the changes made to the codebase contain existing code fragments and thus can be reconstituted using them~\cite{Barr:2014sf,Matias2014Icse}. Many existing works on search-based APR build their techniques upon this hypothesis to reduce the search space of fix ingredients~\cite{Wen2018:cse,TBar2019Issta}. Wen et al. showed that effective fixing ingredients likely exist in the code similar to the target fix location in context~\cite{Wen2018:cse}. Liu et al. highlighted the key role of fixing ingredient retrieval in template-based APR\cite{TBar2019Issta}.
%; the importance of the retrieval of effective fixing ingredients, which often exist near the code with a similar context to the target fix location, has been emphasized multiple times~\cite{Wen2018:cse,TBar2019Issta}. 

In \cref{sub:categories_of_developer_fixes}, we observed that developers often handle Async Wait flakiness by adding or increasing the wait time. With further analysis, we found that developers often select a long wait time, around three times longer on average than the time in the flaky test, for the repair. \name aims to treat an Async Wait flaky test by replacing the wait time in a flaky test with a new time value that can avoid wasting resources while waiting enough for asynchronous calls to finish. Hence, the main objective of the localization phase of \name is to identify the location (i.e., line) to extract these efficient fixing ingredients of \name, which will be the new wait time values. 

\subsubsection{Information-Retrieval based localization}\label{subsub:ir_based_localization}
Given that test flakiness can be considered a specific type of test failure that flakes the outcome, \name adopts an IR-based Fault Localization (IRFL) approach to identify the lines of code where we can extract efficient wait time values. IRFL methods leverage the lexical similarity between the code and bug reports to localize faults in code~\cite{Le:2015:aml,Saha:2013:ase,wen:2016:ase}: \textit{the more similar the code element is to a given bug report, the more likely it is to include a fault.} In this study, we reformulate the idea behind IRFL for our problem and hypothesize: \textit{the more similar a line of code is to the code that causes test flakiness, the more likely the line is to contain an alternative time value (i.e., fix ingredients)}

\name employs a simple IRFL method based on Vector Space Model~\cite{SaltonVSM1975} to compute the lexical similarity between the fix location (i.e., the line that triggers async wait flakiness)\footnote{We assume that developers know where to modify/insert a wait statement. In most cases, locating these lines is relatively easy due to the failure at waiting statement or assertion, as discussed in Section~\ref{sub:categories_of_developer_fixes}} and other lines that contain candidate wait time values; we exclude those without any wait time as out-of-scope. One line before and one line after the fix location, more precisely, the line of the asynchronous call, are processed together to generate a query, providing some context of the flakiness; all other lines become documents to retrieve. We limit our inspection to the lines of test files in the same directory as the flaky test file.

For the preprocessing, we remove JavaScript keywords from the vocabulary, a collection of unique words in documents. Both the query and the documents are then converted into fixed-length weighted vectors. Each weight of this vector corresponds to the term frequency-inverse document frequency (\textit{tf-idf}) of a unique term. The term frequency ($tf$) measures how frequently term $t$ appears within document $d$, while inverse document frequency ($idf$) evaluates how rare the term $t$ is across all documents $D$. Combined, \textit{tf-idf} assesses the importance of the term $t$ in the document $d$: if a rare term frequently appears in a document, this term is likely a keyword of the document. Below equations detail how \textit{tf-idf} is computed for term $t$, document $d$, and a document set $D$.  
\begin{align*}
    \text{\textit{tf(t,d)}} = \frac{f_{t,d}}{\sum_{t' \in d} f_{t',d}} \text{,  }f_{t,d} = \text{the frequency of $t$ within $d$} \\
    \text{\textit{idf(t,D)}} = log{\frac{|D|}{|d \in D: t \in d|}},
    \text{ \textit{tf-idf(t,d,D)}} = \text{\text{\textit{tf(t,d)}}} \cdot \text{\textit{idf(t,D)}} 
\end{align*}
We then compute the cosine similarity between weight vectors of the fix location (i.e., the flaky line) and every candidate line.

\subsubsection{History-based Augmentation}\label{subsub:time_baed_augmentation}

Developers often make similar mistakes more than once, for instance, due to the incomplete understanding of the changes they made~\cite{Kim:2007icse}.
Previous studies of fault localization and defect prediction exploit the re-occurrence of similar failures to improve the accuracy, either directly or indirectly, by mining the project version history~\cite{Foyzur2011Fse,wen:2016:ase,Sohn2021saner}.
\name also mines the version history of a project, inspecting its past changes to augment the pool of candidate locations to find new time values. We assume that developers made or fixed asynchronous flaky tests that failed in a similar context, allowing us to obtain valuable information about the repair from the past. Hence, \name tracks down past changes of the lines identified from the previous IR-based approach and collects past versions of these lines with time values different from those currently have; it also examines the past changes of the \textit{fixed location} since it might undergo (and be treated for) similar asynchronous wait issues before. 
After retrieving all the candidate lines from the present and the past, \name extracts waiting time values, i.e., its fixing ingredients, from these lines; the values smaller than the time used in the flaky test are excluded as they are out of interest. 

\subsection{Tuning and Generation}\label{sub:tuning_and_generation}

The main task of the Tuning and Generation phase is divided into two folds: 1) tune the wait time value and 2) generate a patch with it. \name allows developers to configure the amount of effort spent on the tuning, from only a few trials to upto finding the optimum, depending on the available resources. Under a fast-release cycle, developers can skip this step and go directly to the generation, as shown in \cref{fig:archiecture}.

\name tunes the time value by systematically trying out different values between the original flaky time (the left boundary $B_l$) and the time value obtained from the localization step (the right boundary $B_r$). In the absence of an explicit time delay in the flaky test, such as the tests being flaky for DOM-based causes but addressed by the Time-based fix strategy, $B_l$ is set to be 0. 
For the right boundary $B_r$, \name first sorts the time values in descending order of the similarity score of their source line to the flaky line; if different lines have the same time value, only the one with a higher score is kept. 
It then generates the patch, for instance, via the wait-time replacement, using the time value at the top and validates whether this patch fixes the test flakiness by monitoring 100 test reruns as we did in the preliminary study. \name continues to do this until encountering the time value that resolves the flakiness, i.e., consistently passing 100 runs.  

%Under the static setting, 
By default, \name directly returns $B_r$ and the corresponding fix to the users; we call \name without the extra tuning as \textit{static}, as it does not need any test run until the patch validation. While, in most cases, the candidates from the localization step include at least one valid time value that addresses a flaky test, this is not always ensured: all the found time values can be smaller than the one in the flaky test. \name marks such cases \textit{"non-applicable"}, counting them as failures. 
%\name may fail to include any valid time value through the localization step. 

Under the dynamic mode with tuning, \name basically conducts a binary search for a better value for the wait time with the obtained search boundary ($B_l$, $B_r$).
Algorithm~\ref{alg:dynamic_repair} describes the overall procedure of \name. In general, \name continues to search for a better time value until the time difference between the left and right search boundaries is less than the threshold $thr_{time}$. For each iteration, \name sets the new wait time $wt_{new}$ to be the middle of the current search boundary (line 4). It then attempts to repair the flaky test using $wt_{new}$, checking whether this new test $test_{new}$ is no longer flaky (lines 5,6). If it still is, the left boundary is set to be the current wait time $wt_{new}$; otherwise, \name sets the right boundary to wait time $wt_{new}$ and updates the fixed test to be the corresponding test (line 7-13). 
The search ends when the search range becomes smaller than a given threshold $thr_{time}$.

\begin{algorithm}[t]
\SetAlgoLined
\SetKwInOut{Input}{input}\SetKwInOut{Output}{output}
\Input{project $P$, flaky test $test_{flaky}$, 
    the left and right boundary, $B_l$ and $B_r$, 
    time threshold $thr_{time}$, 
    the number of test executions $N_{exec}$
}
\Output{the repaired test $test_{fixed}$}%

$test_{fixed} \gets$ repair($test_{flaky}$, $B_r$)

$\Delta_{time} \gets B_r - B_l$ 

\While {$\Delta_{time} > thr_{time}$} { 
    
    $wt_{new} \gets \frac{B_l + B_r}{2}$
    
    $test_{new} \gets$ repair($test_{fixed}$, $wt_{new}$)

    $isFlaky \gets$ executeAndCheck($P$, $test_{new}$, $N_{exec}$)
        
    \eIf {$isFlaky$}{
        $\Delta_{time} \gets B_r - wt_{new}$ 

        $B_l \gets wt_{new}$
    }{
        $\Delta_{time} \gets wt_{new} - B_l$ 

        $B_r \gets wt_{new}$

        $test_{fixed} \gets test_{new}$
    }
}

\Return $test_{fixed}$

\caption{Dynamic Repair with Tuning}\label{alg:dynamic_repair}
\end{algorithm}

\subsubsection{Repair}\label{subsub:Repair}

\begin{figure}[h]
  \centering
  \includegraphics[width=0.9\linewidth]{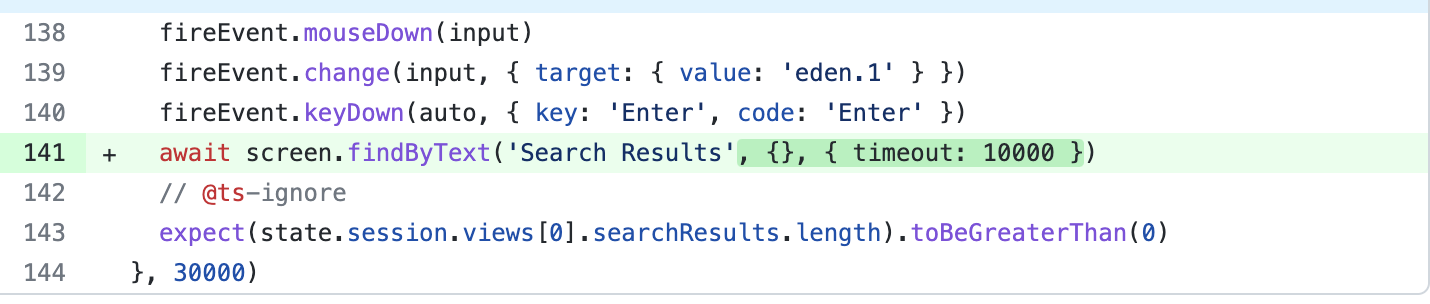}
  \vspace{-0.2em}
  \caption{The repair case without an explicit timeout but with the timeout option in an asynchronous call.} 
  \label{fig:repair-wait-after}
\end{figure}

\begin{figure}[h]
  \centering
  \includegraphics[width=0.9\linewidth]{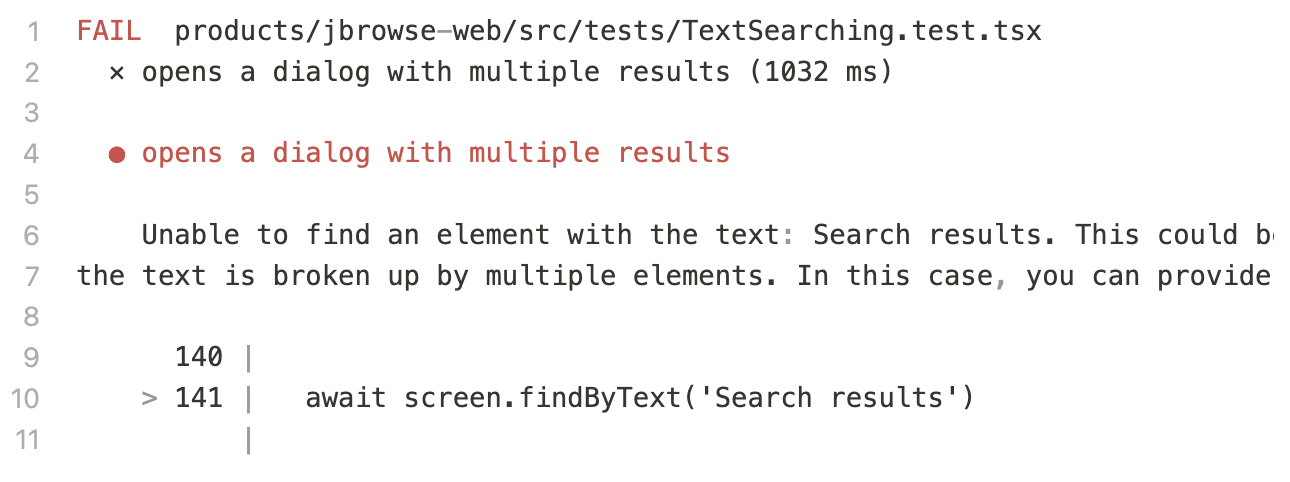}
  \vspace{-0.2em}
  \caption{Error message with location info of the repair case.}
  \label{fig:repair-error}
\end{figure}

\name attempts to handle the flaky test using a new wait time value with the \textit{repair} function in Algorithm~\ref{alg:dynamic_repair}. 
In many cases, the repair procedure is straightforward: it replaces the wait time value of an asynchronous call in the flaky test with the new one, as shown in \cref{fig:time-based}. 
For the cases without an explicit wait time, \name adds the obtained wait time as an additional timeout to the asynchronous call to a DOM element, as in \cref{fig:repair-wait-after} from project Jbrowse-components, or inserts an additional wait function after the call. 
Here, we assume developers to provide the fix locations, i.e., the location of an asynchronous call. In our study, we take the last statement in the stack trace;\footnote{The error messages of two cases in our dataset failed to identify fix-location (i.e., line) due to being flaky by interactions of multiple asynchronous calls. For these and only these, we used the developers' fix locations.} as shown in \cref{fig:repair-error}, this information can be easily obtained from Async Wait flaky tests in many cases.

\vspace{-1em}

\section{Experimental Settings}\label{sec:experimental_settiings}

\subsection{Research Questions}\label{sub:research_questions}

With the aim of evaluating the performance and usefulness of \name, we ask the following three research questions. % in fixing asynchronous wait flakiness issues. 

\begin{description}
\item[RQ1]\textbf{Effectiveness:} How effective is \name in finding the right  wait time to repair Async Wait flakiness? 
\label{subsub:rq1}
\end{description}

We evaluate the effectiveness of \name by comparing its generated wait time with the time used in developer-written patches. We further compared the execution time of the tests repaired by \name and by developers to study whether the improvement in wait time can be beneficial in practice. 

\name has two modes, static ($\name^S$) and dynamic ($\name^D$). \name runs in static mode by default, and users can choose to refine the found wait time values via dynamic tuning, as illustrated in \cref{fig:archiecture}. We evaluate the performance of \name in each mode, investigating whether it can statically find time values that can address Async Wait flakiness and how the dynamic tuning can further reduce the execution time.

Section~\ref{subsub:Repair} explains that some flaky tests are more complex to repair than others. Hence, we assign three levels to 31 flaky tests according to their repair difficulty: the higher the level is, the more challenging to repair. \textit{Level 0} includes tests that can be repaired by increasing the existing wait time. \textit{Level 1} means that while tests do not have any explicit timeout call, they can be addressed by adding the wait time before or to a single asynchronous wait call to a DOM element. Lastly, tests that belong to \textit{Level 2} are those where interactions of multiple statements result in an asynchronous wait. We also conduct our evaluation on these three groups for the following research questions.

\begin{description}
\item[RQ2] \textbf{Efficiency:} %How much effort is required for \name to repair Async Wait flakiness? 
How efficient is \name at repairing Async Wait flaky tests?
\label{subsub:rq3}
\end{description}

\name requires a test rerun to validate a generated patch. The patch validation takes place under two circumstances: 1) after the localization step to select the most likely time value from the candidate list and 2) for every search iteration of the dynamic tuning. RQ1 focuses on evaluating the effectiveness of \name without ($\name^S$) and with the tuning ($\name^D$). In RQ2, we investigate the cost of obtaining the observed performance by inspecting the number of test reruns in \name. 

While the dynamic tuning process of \name optimizes the time values from the previous localization, it may not always be feasible, especially for projects under the fast release cycle, as the test needs to be rerun for every optimization attempt. To better understand the trade-off between the effectiveness and efficiency of this tuning, we further study the performance between $\name^S$ and $\name^D$, configuring \name to try out a few shots of tuning until meeting the first valid time value that resolves the flakiness with a shorter wait time. We call this tuning configuration \textbf{\textit{first-match}}, as the optimization ends immediately after the first successful refinement; we refer $\name^D$ with the first-match tuning as $\name^D_{FM}$. 

In the end, we compare the number of test reruns between these three configurations in terms of efficiency; to get a more tangible view, we further relate the results to the test execution time. We differentiate the number of reruns during the dynamic tuning, as it is an additional cost. We refer to the total number of test reruns as $n_{rerun}$ and those during the tuning as $n_{iter}$.

\begin{description}
\item [RQ3]\textbf{Augmentation Contribution:} How does the history-based augmentation contribute to effectiveness of \name?
\label{subsub:rq2}
\end{description}

To confirm that the augmentation step using the past change history contributes to obtaining a better waiting time, we compare the performance of \name with and without the augmentation in both static and dynamic modes. More specifically, we compute the difference between the waiting time and the execution time acquired by \name with and without the augmentation, inspecting whether the augmentation allows \name to find a more efficient latency value to resolve an asynchronous issue. The current codebase may not contain any valid time value, resulting in \name being \textit{non-applicable}, as described in \cref{sub:tuning_and_generation}. Hence, we further compute the Repair Rate ($RR$) of \name, which is the ratio of the number of cases where \name successfully finds a valid time value, i.e., applicable, over the total number of studied cases, including the failed ones, i.e., non-applicable.
%We further investigate whether there is a specific scenario in which this augmentation can be particularly contributing. 

\subsection{Baseline}\label{sub:baseline} 
\name addresses Async Wait flakiness by suggesting a more efficient wait time for asynchronous calls. To evaluate its performance, we employ two baselines: 1) using the original wait time in developer-written fixes and 2) using the refined version of this developer-chosen wait time via FaTB~\cite{Lam2020a}. 

FaTB is a post-processing approach that further refines the wait time used by developers to address Async Wait flakiness, aiming to reduce test execution time while retaining the decreased flake rate of developer fixes. It systematically tries out different time values between the original flaky time and the time chosen by developers to address the flakiness.  
FaTB and the optimization step of \name share the same search procedure for the refinement, checking the middle of two boundary values to narrow down the search space. As a result, letting the refinement to the end will result in FaTB and $\name^D$ converging at the same value close to the optimal. 
Hence, we implement FaTB to stop at the first successful refinement, comparing it mainly with $\name^D$ under first-match configuration ($\name^D_{FM}$). 
As the goal here is to evaluate how efficient the wait time suggested by \name is than the time selected by developers, we change FaTB to use the resolution of flakiness instead of the flake rate, configuring its tolerance of flaky test failures as zero. 
We refer to this implementation of FaTB as FaTB$_{FM}$ since it terminates at the first refinement similar to \textit{first-match} configuration of \name. 

\subsection{Version History Collection}\label{sub:version_history_collection}

\name augments the candidates to extract time values by further inspecting their past versions, assuming that developers often make similar mistakes repeatedly. To obtain the past versions of lines, we tracked down their past changes using Git v.2.37.0~\cite{chacon2014Git} and difflib module of Python 3.9.15\footnote{SequenceMatcher determines the line deletion, insertion, and modification}.

\subsection{Configuration}\label{sub:algorithm_configuration}

\name allows users to configure the effort spent on the tuning via the time threshold $thr_{time}$, the minimum search range. We set this threshold to 5 ms in our experiment, investigating how much test execution time \name can save at most if there are enough resources for tuning. 

\subsection{Implementation and Environment}

\name is implemented in JavaScript and runs on Node.js v17.5.0 and NPM v8.4.1. We ran all experiments on the machine equipped with Apple M1 Max and 32G RAM.

\section{Results}\label{sec:result_analysis}

\subsection{RQ1. Effectiveness}\label{sub:rq1_result}
 
\cref{tab:rq1_table_1} shows the effectiveness of \name in both static (\name$^S$) and dynamic ($\name^D$) modes compared to the developer-written fixes ($dev$) in the wait time and the test execution time.
Overall, \name successfully finds shorter wait time values than those chosen by developers. Under the static mode without the tuning, \name reduces the average test execution time by 11.1\% (186.6 ms) per test compared to the initial test execution time; the wait time decreases by an average of 47.4\%, from 6528.1 ms to 3433.2 ms.
For each repair difficulty level, \name shows different performance. For flaky tests with difficulty levels 0 and 1, \name saves the average execution time and wait time by 26.0\% (367.6 ms) and 32.6\% (418.5 ms) (Level 0) and by 1.3\% (23.7 ms) and 48.8\% (5625.0 ms) (Level 1). Most of the flaky tests in Level 1 are handled by adding a timeout to an asynchronous call to a DOM element. Therefore, we suspect that the notable difference in average improvement between wait time and execution time observed at level 1 is related to function calls that often terminate before a given timeout during test reruns by satisfying DOM-related conditions in the same call. For those in level 2, the execution time and wait time decrease by 15.4\% (312.5 ms) and 45.5\% (250 ms).
With dynamic tuning, $\name^D$ further improves the performance by finding a wait time close to the optimum, reducing test execution time by an average of 20.2\% compared to the developer fixes.

\begin{figure*}[t!]
\vspace{-0.5em}
    \centering
    \begin{subfigure}{0.475\textwidth}
        \includegraphics[width=\textwidth, trim=2mm 0mm 0mm 0mm, clip]{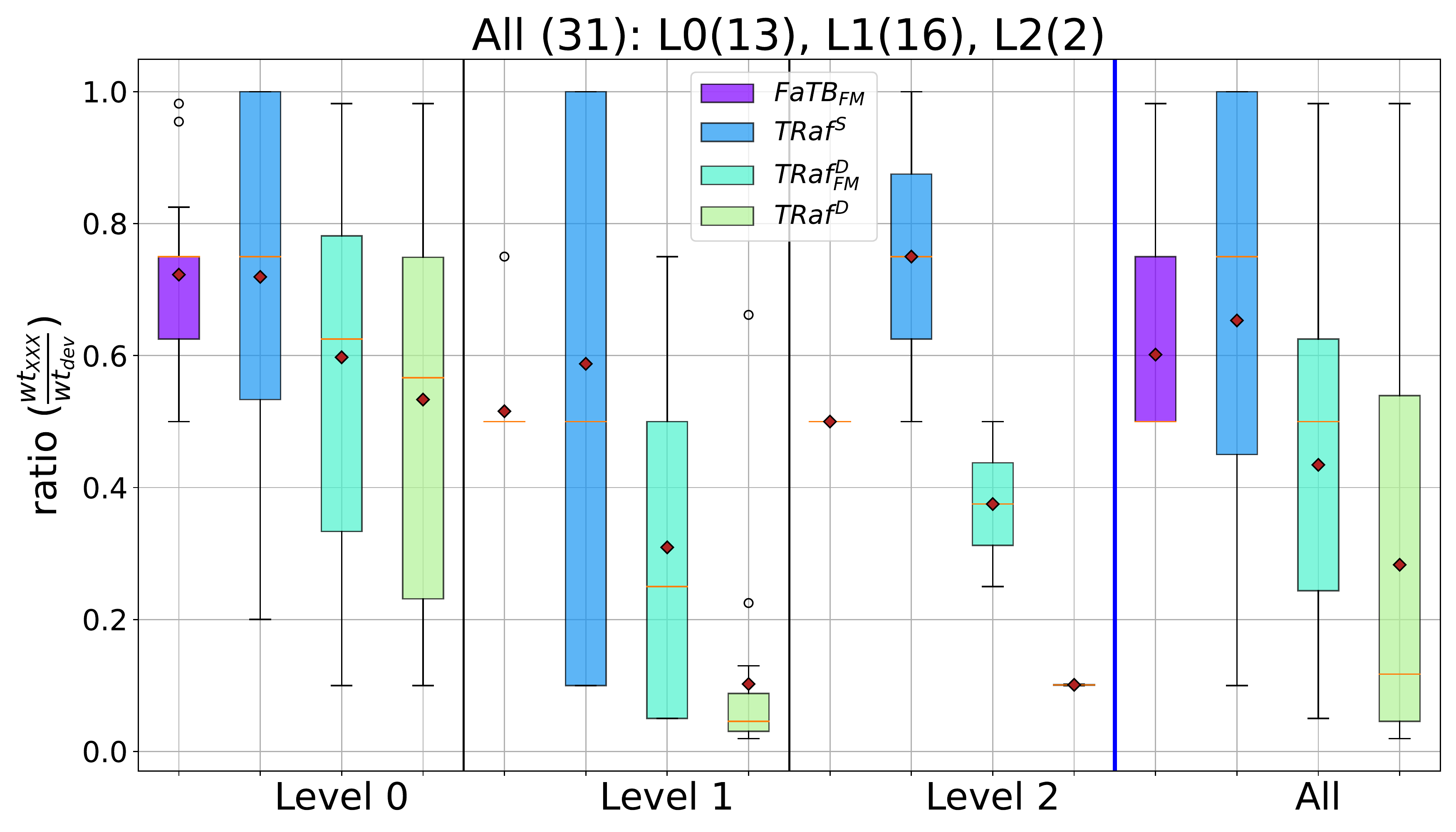} 
        \caption{
        Waiting Time Comparison
        \label{fig:rq1_wt}}
    \end{subfigure} 
    \begin{subfigure}{0.475\textwidth}
        \includegraphics[width=\textwidth, trim=2mm 0mm 0mm 0mm, clip]{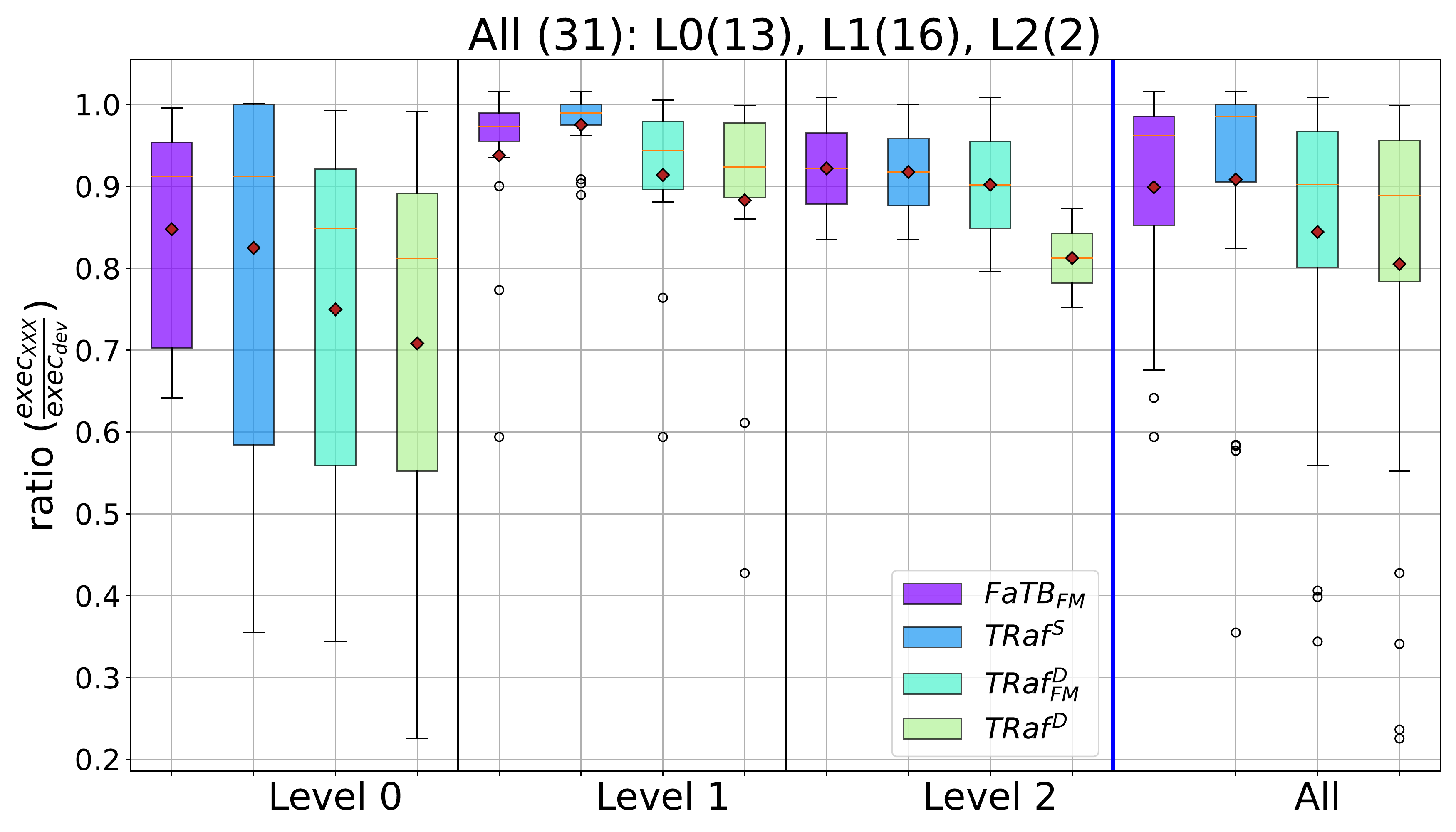}
        \caption{
        Execution Time Comparison
        \label{fig:rq1_et}}
    \end{subfigure}
\vspace{-0.5em}
    \caption{Effectiveness. Comparison between the variations of \name and the developer-fix-time by computing the ratio (y-axis). 
    The first plot refers to the baseline FaTB. $L0$, $L1$, and $L2$ denote difficulty levels and $All$ refers to including all flaky tests.} 
    \vspace{-0.05em}
    \label{fig:ratio_comparison}
\end{figure*}

\cref{fig:ratio_comparison} summarizes the effectiveness of \name compared to the developer-written patches for Async Wait flaky tests; the y-axes of \cref{fig:rq1_wt,fig:rq1_et} are the ratios of the new wait time and the respective execution time to the time selected by developers and the corresponding execution time, respectively: the smaller the ratio is, the more effective \name is.
Overall, \name reduces the wait time and, thus, the execution time for most of the flaky tests, as summarized in the last three plots in these figures (\textit{All}).\footnote{For flaky tests in Levels 1 and 2, the execution time sometimes degrades, i.e., the ratio over 1, due to early termination and taking the average. However, these are rare cases, and \name still finds proper timeouts for these cases.} 
Here, the individual improvements in test execution time are less consistent than in wait time. We speculate that this may relate to the difference in the testing environment, for example, resulting in test runs that initially fail with a given latency value to pass more frequently; as the test execution time values in \cref{tab:rq1_table_1} are the average of 100 test runs, the impact of using better time values may be dimmed by the passing runs. Nevertheless, in most of the studied flaky tests, \name discovers more efficient wait time values than developers, with or without the tuning, showing the existing and potential benefits of efficiently handling Async wait flakiness.

The wait time and the execution time are reported in milliseconds. Hence, the reported improvement could be perceived as trivial with the absolute values alone. However, we would like to emphasize that the improvement here is per test execution. Given the prevalence of flaky tests in a development environment that requires intensive testing regularly, e.g., a project under continuous integration, we posit that the collective reduction of these small improvements can save a substantial amount of future expenses of testing; \textbf{\cref{sec:discussion}}, i.e., \textit{Discussion}, will detail the potential benefits in practice.

\textbf{Answer to RQ1:} 
\name can effectively repair the time-based Async Wait flakiness by finding a better wait time that decreases the test execution time by 20.2\% than developer-written fixes while still resolving the flakiness. 

\begin{table*}[ht]  
\centering
\caption{Comparison of the average wait and test execution time of \name, developers' fixes, and FaTB. n$_{t}$ and n$_{dm}$ are the number of Time-based and DOM-based flaky tests, respectively;
0, 1, and 2 are flaky test difficulty levels. \name finds a shorter wait time than the developers' time values ($dev$), in static $\name^{S}$ and dynamic $\name^{D}$ modes, while resolving the flakiness. 
}\label{tab:rq1_table_1}
%\vspace{-0.2em}
\scalebox{0.9}{
\begin{tabular}{lr|rrrrr|rrrrr}
    \toprule

     &  & \multicolumn{5}{c|}{Wait Time (ms)} & \multicolumn{5}{c}{Execution Time (ms)} \\
    Level & n (n$_{t}$/n$_{dm}$) & $dev$ & FaTB$_{FM}$ & $\name^{S}$ & $\name^{D}$ & $\name^D_{FM}$ & dev & FaTB$_{FM}$ & $\name^{S}$ & $\name^{D}$ &  $\name^D_{FM}$  \\
    \midrule
    0 & 13 (13/0) & 1282.3 & 884.8 & 863.8 & 609.7 & 714.6 & 1413.2 & 1120.2 & 1045.6 & 877.6 & 945.0\\
    1 & 16 (1/15) & 11537.5 & 5815.6 & 5912.5 & 646.1 & 3003.1 & 1844.8 & 1755.8 & 1821.1 & 1686.4 & 1727.9\\
    2 & 2 (1/1) & 550.0 & 275.0 & 300.0 & 56.0 & 150.0 & 2022.8 & 1711.3 & 1710.3 & 1536.0 & 1635.8\\
    
    \midrule
    total & 31 (15/16) & 6528.1 & 3390.4 & 3433.2 & 592.8 & 1859.4 & 1675.3 & 1486.4 & 1488.7 & 1337.5 & 1393.7\\
    \bottomrule
\end{tabular}
}
\end{table*}

\subsection{RQ2. Efficiency}\label{sub:rq2_result}

\cref{tab:rq2_table_1} presents the total number of test reruns in \name ($n_{rerun}$), i.e., from the localization to the repair and validation, and only during the dynamic optimization ($n_{iter}$). For the baseline FaTB$_{FM}$, both $n_{iter}$ and $n_{rerun}$ refer to the number of test reruns during the refinement, plus one, since there is no need for localization in FaTB by having the developer’s time beforehand; one additional run is to check the validity of time values in developer fixes.  

Column $\name^S$ in \cref{tab:rq2_table_1} shows that the localization module of \name often locates the line with a valid time value to treat test flakiness at the top of the candidate list. In fact, for 25 out of 31 (81\%) flaky tests, \name addresses the flakiness using the time value at the top without inspecting further; the average ranking of the line with a proper waiting time is 1.2.
%that the localization module of \name often ranks the line that has  a valid time value to resolve test flakiness at the top of the candidate list.

Regarding the dynamic tuning, \name successfully refines a given time value within two trials, as shown in $n_{iter}$ in column $\name^D_{FM}$ of \cref{tab:rq2_table_1}. The performance with these refinements is reported in the last columns, $\name^D_{FM}$, in \cref{tab:rq1_table_1}. As expected, $\name^D_{FM}$ falls between $\name^S$ and $\name^D$ in effectiveness and expense, showing the trade-offs between these criteria. Nevertheless, the average execution time per test of $\name^D_{FM}$ is only around 56 ms (4.0\%) longer than the execution time of tests using fully optimized time values from $\name^D$; this shows the additional improvement that a few trials of tuning can easily obtain. 

In addition to the developer fixes, we further compare \name, specifically $\name^D_{FM}$, with FaTB$_{FM}$, i.e., the counterpart of $\name^D_{FM}$ in the aspect of using the time value chosen by developers instead of the one obtained by comparing with the current and past codebase.
The columns FaTB$_{FM}$ in \cref{tab:rq1_table_1,tab:rq2_table_1} present the performance and the cost of running FaTB$_{FM}$;  
In general, $\name^D_{FM}$ outperforms FaTB$_{FM}$ while going through a similar number of iterations during the optimization; for instance, in \cref{fig:ratio_comparison}, both the median and the mean of $\name^D_{FM}$ (i.e., the third plots) are often smaller than those of FaTB$_{FM}$ (i.e., the first plots). 
From these observations, we argue that \name can efficiently propose to developers in advance an adequate wait time to mitigate asynchronous flakiness while minimizing execution time.

\textbf{Answer to RQ2:} 
\name can efficiently discover adequate wait time values to handle Async Wait flakiness issues without requiring many test reruns (i.e., 81\% by the first trial), even with the additional tuning (e.g., at most two trials for first refinement). The comparison between \name and time values suggested by developers, both the original and the refined, further supports the usefulness of \name.

\subsection{RQ3. Augmentation Contribution}\label{sub:rq3_result}

\cref{tab:rq3_table_1} shows the impact of using the history-based augmentation in \name.
For 31 flaky tests that we studied, \name failed to find any valid latency value for three cases\footnote{These three case projects have a small number of test files, and only the current file has the wait time statement.} without the aid of the augmentation. Since \name cannot be applied to these cases, we consider them failures. 
Hence, by comparing the repair rate ($RR$) alone, which is 90.3\% for \name without the augmentation ($\name_{woa}$) and 100\% for \name with the augmentation, we posit that the augmentation step contributes positively to the performance of \name. 
Nevertheless, in order to confirm that the augmentation indeed helps to find a more efficient timeout, we further compare the wait time and the execution time of \name with and without the augmentation by computing their differences, as shown in \cref{tab:rq3_table_1}; the difference values reported in the table were calculated for 28 cases (i.e., 90.3\%) where \name found a valid time value even without the augmentation. 

Overall, with the augmentation, \name often finds a smaller latency value that still resolves the flakiness issue while reducing the test execution time than without the augmentation. For instance, for flaky tests from Level 0, $\name^S$ selected a timeout on average 115.4 (ms) (11.8\%) smaller than the ones chosen without the augmentation. 
Similarly, in $\name^D_{FM}$, the timeout value is, on average, 37.6 (ms) (5.0\%) smaller than the timeout without the augmentation; the execution time reduces by 12.9 (ms) with the augmentation. While the augmentation allows \name to generate a more efficient timeout value in most cases, it sometimes finds a less efficient one, increasing the execution time compared to without the augmentation. Nonetheless, for the majority of the studied tests, the augmentation aids \name in finding a better waiting time, as \cref{tab:rq3_table_1} demonstrates, and, more importantly, increases the repair rate, reaching 100\%.  

\textbf{Answer to RQ3:} The augmentation step improves the effectiveness of \name, allowing it to reach a 100\% repair rate under 100 reruns. It also allows \name to find a more efficient wait time value that further reduces the test execution time. 

\begin{table}[ht] 
\caption{
Comparison of the average number of test reruns ($n_{rerun}$, $n_{iter}$) and the repair rate ($RR$). $\name^S$ always has zero for $n_{iter}$, as there is no dynamic tuning. 
$\text{Perc}_{1}$ is the percentage of cases where \name ranks a proper time value at the top and is reported only for $\name^S$, as the tuning takes $\name^S$'s output as input. %\name addresses all Async Wait tests successfully ($RR$).
}\label{tab:rq2_table_1}
\scalebox{0.8}{
\begin{tabular}{lr|r|rrrr}
    \toprule
     &  & & \multicolumn{4}{c}{$n_{iter}$ / $n_{rerun}$ (\text{Perc}$_{1}$ (\%)) }  \\
    Lev & n (n$_{t}$/n$_{dm}$) & $RR$ (\%) & FaTB$_{FM}$ & $\name^S$ & $\name^D$ & $\name^D_{FM}$ \\ 

    \midrule
    0 & 13 (13/0) & 100.0 & 2.2 / 2.2 & 0.0 / 1.2 (85\%) & 5.1 / 6.3  & 1.7 / 2.9 \\
    1 & 16 (1/15) & 100.0 & 2.7 / 2.7  & 0.0 / 1.3 (75\%) & 7.1 / 8.4 & 1.4 / 2.7 \\
    2 & 2 (1/1) & 100.0  & 2.0 / 2.0 & 0.0 / 1.0 (100\%) & 5.5 / 6.5 & 1.0 / 2.0 \\
    \midrule
    total & 31 (15/16) & 100.0 & 2.5 / 2.5 & 0.0 / 1.2 (81\%) & 6.2 / 7.4 & 1.5 / 2.7 \\
    \bottomrule
\end{tabular}}
\end{table}

\begin{table}[ht] 
\centering
    \caption{
    Comparison of repair rate (RR) and average wait time (WT) and execution time (ET) between static and dynamic \name, with and without augmentation. The differences in WT (dff$_{WT}$) and ET (dff$_{ET}$) are in $ms$. Negative means that the one with the augmentation outperforms the one without it. }\label{tab:rq3_table_1}
    \scalebox{0.78}{
    \begin{tabular}{lr|rr|r|rr|r}
        \toprule
        & & \multicolumn{2}{c|}{$\name^S$ vs $\name^S_{woa}$} & $\name^S_{woa}$ & \multicolumn{2}{c|}{$\name^D$ vs $\name^D_{FM,woa}$} & $\name^D_{FM,woa}$ \\
        $Lev$ & n & dff$_{WT}$ & dff$_{ET}$ & $RR$ (\%) & dff$_{WT}$ & dff$_{ET}$ & $RR$ (\%)  \\

       \midrule
        0 & 13 & -115.4 & -48.9  & 100\%     & -37.6 & -12.9  & 100\%  \\
        1 & 14 & 0      & 0     & 87.5\%    &  0    & 0      & 87.5\% \\
        2 & 1  & 0      & 0     & 50\%      &  0    & 0     & 50\%   \\
        \midrule
    total & 28 & -53.5  & -22.7  & 90.3\%    & -17.4 & -6    & 90.3\% \\
        \bottomrule
        
    \end{tabular}
    }
    %\vspace{-1.0em}
    \end{table}
\section{Discussion}\label{sec:discussion}

\noindent\textbf{The Prevalence of Async Wait Flaky Test and the Impact of Addressing them.}
Asynchronous wait flaky test, shortly Async Wait test, is one of the most common types of flaky tests observed in open-source and proprietary projects~\cite{Lam2020a,Luo2014,Romano2021ICSE,Parry2021Tosem,Hashemi2022icsme,malm2020automated}. Luo et al. studied 201 flaky test-fixing commits from 51 open-source projects to identify the common root causes and fix strategies adopted in practice. Out of these 201 commits, 74 (36.8\%) were related to fixing Async Wait flaky tests~\cite{Luo2014}; among these 74 flaky tests, 39 adopted a fixing strategy of adjusting the wait time or timeout. Similar statistics were observed in a study of six large-scale projects at Microsoft~\cite{Lam2020a}: out of 134 fixed flaky tests studied by the authors, 87 fixes were related to asynchronous method calls, and 31\% of them involved increasing the wait time, revealing developers’ preference toward a simple time-based fix strategy. In this paper, we studied flaky tests in web front-end testing. Recently, Romano et al. conducted an empirical study of flaky tests in UI testing, reaffirming the previous findings about Async Wait flakiness and how developers handle it~\cite{Romano2021ICSE}.

The findings on Async Wait flaky tests from these prior empirical studies imply that although the time-based fix strategy of increasing or adding a timeout cannot completely remove flakiness, it is one of the most adopted by developers for Async Wait tests. Indeed, we observe a similar trend in our preliminary study (\cref{sec:subject_study}). As the time-based strategy does not need developers to identify the root cause of asynchronous flakiness, it can be employed as a hotfix to alleviate the flakiness and ensure the continuous delivery of software. Nevertheless, this usability often comes at the cost of increased test execution time. The main idea of this time-based strategy is to wait long enough for an asynchronous call to complete. Hence, without knowing what is \textit{"enough"}, which is often the case, it is challenging to choose a proper timeout to resolve the flakiness. As a result, developers tend to set a large new timeout, increasing the test expenses. Given that regression testing has become a standard development practice, this increase in cost, however small it is per test, may easily come back as a huge burden in testing. \name tackles this issue by finding an efficient time to wait for an asynchronous call, using code similarity and past changes, assuming that the hints for the right time value exist in the same codebase of present and past near the code with a similar context to a fix-location.

The experimental results with 31 Async Wait flaky tests show that the time values from the lines with similar context can address the flakiness more efficiently than the time values in the developer’s fixes. \name achieved this performance almost without any test rerun involved, successfully ranking the proper wait time at the top of its candidate list in most cases (25 out of 31 cases); within a few trials of dynamic refinement, i.e., at most two, \name further refined the localized values. This low expense of \name demonstrates its potential to be used on-the-fly to handle Async Wait flakiness. Considering the prevalence of Async Wait flakiness in practice, we believe \name can support delivering reliable software continuously.

\section{Threats to Validity}\label{sec:threats_to_validity}

%This work is subject to many of the common threats typically found in empirical studies. In this section, we focus on the issues that are more specific to our findings.

%The threat to internal validity comes from the validation process of the evaluation results, such as the dataset we used and the re-run of experiments. 
%According to our findings, flaky tests exhibit some practices in terms of reproducibility and execution efficiency. The efficiency of finding flaky tests can also be affected by repeated executions, particularly if the test cases are relatively expensive to execute. Therefore, to balance the impact of the number of runs and execution time, we choose to run each test 20 times.

The threat to internal validity relates to the validity of our evaluation, i.e., whether the evaluation can support what we claim. 
To avoid introducing biases into the studied flaky tests in web testing, we systematically collected 49 flaky tests from 26 open-source projects containing the most employed web application testing frameworks. 
We evaluate the generated repair by monitoring the pass-and-fail state of a given test across its repeated runs.  
Previous studies have shown that 100 reruns are typically enough to determine whether a test is flaky or not~\cite{Lam2019RootCausing, Dutta2020}. Hence, to ensure the validity of \name’s fix for Async Wait flaky tests, we rerun tests 100 times.  

%According to our findings, flaky tests exhibit some practices in terms of reproducibility and execution efficiency. The efficiency of finding flaky tests can also be affected by repeated executions, particularly if the test cases are relatively expensive to execute. Therefore, to balance the impact of the number of runs and execution time, we choose to run each test 20 times.
External validity refers to the factors that impact the generalizability of the conclusion. While our target is restricted to projects that can run locally, we improved generalizability by including diverse projects with different testing frameworks. This study focuses on examining JavaScript-written projects, as it is the dominant language in web development. 
% beforechange In this study, we only studied projects written in JavaScript, which is the most dominant language in web development. 
%We restricted the use of projects from open-source projects that could be run in our experimental environment to obtain reproducible test cases and to verify the effect of the fixes, due to the platform dependencies, and the inaccessibility of some test suites. 
%Our conclusions may be limited by the choice of language(JavaScript), and the size of the dataset, as well as factors in the studied projects that we failed to run. 

The threats to construct validity include our manual data collection and classification checks. We collected the commits related to the flaky test, filtered out irrelevant commits, and manually inspected the commits to verify their relevance to the async wait flaky test. Furthermore, to minimize errors, two authors independently performed three rounds of inspections for categorizing async wait flaky tests, considering commit information, source code, error messages, and fixed code, and derived a classification from the discussion.

\section{Related work}\label{sec:related_work}

Flaky tests exhibit non-deterministic and unpredictable behaviors that require to be addressed differently. Many prior studies have attempted to categorize flaky tests based on their prevalence in software testing~\cite{Luo2014,Gruber2021,Hashemi2022icsme,Parry2021Tosem,Lam2019RootCausing,Rahman2018,Romano2021ICSE}. 
Luo et al. conducted an extensive study on flaky tests, studying 201 commits from 51 open-source projects, and investigated what triggers flaky tests, how they manifest, and how they can be fixed and grouped based on their findings~\cite{Luo2014}.
Lam et al. studied the logs of flaky tests in an industrial setting and 
%found that while the number of builds failed by flaky tests might be substantial
%, the number of distinct tests is rather limited~\cite{Lam2019RootCausing}. Thus, they 
and proposed RootFinder, a framework to identify the root cause of a given flaky test by analyzing the logs of failing and passing tests. A recent study with proprietary Microsoft projects confirmed that existing findings on flaky tests from open-source projects, including common causes and fix strategies, generalize to projects with a different level of accessibility~\cite{Lam2020a}; in this work, the authors observed that developers often adapt the timeout to alleviate flakiness due to asynchronous calls.  

Web testing involves various factors that are inherently unpredictable, such as user interactions and network delays, and are thereby susceptible to flakiness~\cite{Romano2021ICSE,Hashemi2022icsme}. Romano et al. performed an empirical study on flaky tests in UI-testing in web and Android environments~\cite{Romano2021ICSE}; the study revealed that flaky tests in UI-testing also have similar categories of root causes and fix strategies in traditional unit testing. 
All these prior studies, regardless of the different subjects and environments, agree that flaky tests with different causes require different fixing strategies. In this study, we focus on the flaky tests caused by Asynchronous waits, exploring their more detailed root causes and common fix strategies of developers in web front-end testing.

Previous empirical studies on flaky tests confirmed that developers addressed flaky tests differently depending on their causes. Hence, existing works on automated flaky test repair focused on repairing a certain type of flaky tests~\cite{Dutta2021fseFLEX,shi2019ifixflakies,olianas2022sleepreplacer,Lam2020a,Shi2019Mitigating,Li2022icseODRepair}. 
Shi et al. proposed iFixFlakies, a tool to automatically fix Test-Order Dependency flaky tests that non-deterministically fail and pass depending on the test execution order~\cite{shi2019ifixflakies}. iFixFlakies resolves the order-dependent flaky tests by calling helper functions that recover shared states before pollution.
Recently, Li et al. introduced ODRepair, which addresses the weakness in iFixFlakies that assumes the existence of helper functions~\cite{Li2022icseODRepair}. ODRepair generates the code to clean up polluted states of tests rather than looking for helpers.  
Async Wait flaky tests are one of the most common flaky tests in practice~\cite{Parry2021Tosem}. While relatively small in number compared to repair works on other types of flaky tests, recent studies demonstrated the potential of automated repair of Async Wait test flakiness. Lam et al. proposed the FaTB that dynamically improves the wait time selected by developers to reduce the test execution time~\cite{Lam2020a}. 
In the area of web testing, Olianas et al. introduced an approach that automatically repairs Async Wait flaky tests by replacing thread sleeps with explicit waits in an E2E Selenium WebDriver test suite~\cite{olianas2022sleepreplacer}.
%Outside of traditional types of flaky tests, Dutta et al. presented FLEX, the first technique for automatically fixing tests that are flaky due to algorithmic randomness in ML algorithms~\cite{dutta2021flex}. 
Our approach \name shares similarities with previous studies on Async Wait flaky test repair.
However, \name differs from the existing methods in that it does not need to know the developers' wait time in advance and can efficiently suggest a new time for waiting by adapting the plastic surgery hypothesis. By simply adapting the wait time for repair instead of replacing it with an explicit wait for a certain element, \name requires far less time to address the flakiness. % In other words, \name can complement these approaches without much expense, as it can be run statically and without user-related inputs. 
\section{Conclusion}\label{sec:conclusion}
In this paper, we conducted an empirical study on Async Wait flakiness in web front-end testing. We investigated 49 reproducible Async Wait flaky tests collected from 26 JavaScript web projects. We divided these tests according to the way developers fixed them, confirming that adding or increasing the wait time is the most common repair practice in web front-end applications for this type of flaky tests. Hence, we proposed \name, a novel time-based repair framework that automatically addresses Async Wait flakiness in web front-end testing. \name leverages code similarity to flaky lines and the version history of code under test to obtain efficient time values to alleviate test flakiness while minimizing the execution time. 
Our empirical analysis shows that \name can outperform fixes written by developers by finding more efficient wait time values that reduce the test execution time by 11.1\% and by 20.2\% through additional dynamic tuning of the found wait time while still reducing test flakiness.

\bibliographystyle{IEEEtran}
\bibliography{newref,references}

\end{document}